\documentclass[aps,prl,numerical,amsmath,twocolumn,amssymb]{revtex4-1}
\usepackage{graphicx,epsfig,epsf}
\usepackage{dcolumn}
\usepackage{bm}
\usepackage{amsmath}
\usepackage{amssymb}
\usepackage{dsfont}
\usepackage{graphicx}
\usepackage{color}
\usepackage[normalem]{ulem}

\begin{document}
\newcommand{\ri}{{\rm i}}
\newcommand{\re}{{\rm e}}
\newcommand{\bb}{{\bf b}}
\newcommand{\bc}{{\bf c}}
\newcommand{\bx}{{\bf x}}
\newcommand{\bz}{{\bf z}}
\newcommand{\by}{{\bf y}}
\newcommand{\bv}{{\bf v}}
\newcommand{\bd}{{\bf d}}
\newcommand{\br}{{\bf r}}
\newcommand{\bk}{{\bf k}}
\newcommand{\bA}{{\bf A}}
\newcommand{\bE}{{\bf E}}
\newcommand{\bF}{{\bf F}}
\newcommand{\bR}{{\bf R}}
\newcommand{\bM}{{\bf M}}
\newcommand{\bn}{{\bf n}}
\newcommand{\bs}{{\bf s}}
\newcommand{\tbs}{\tilde{\bf s}}
\newcommand{\rSi}{{\rm Si}}
\newcommand{\beps}{\mbox{\boldmath{$\epsilon$}}}
\newcommand{\bthe}{\mbox{\boldmath{$\theta$}}}
\newcommand{\blam}{\mbox{\boldmath{$\lambda$}}}
\newcommand{\rg}{{\rm g}}
\newcommand{\xmax}{x_{\rm max}}
\newcommand{\ra}{{\rm a}}
\newcommand{\rx}{{\rm x}}
\newcommand{\rs}{{\rm s}}
\newcommand{\rP}{{\rm P}}
\newcommand{\up}{\uparrow}
\newcommand{\down}{\downarrow}
\newcommand{\hc}{H_{\rm cond}}
\newcommand{\kb}{k_{\rm B}}
\newcommand{\cI}{{\cal I}}
\newcommand{\tit}{\tilde{t}}
\newcommand{\cE}{{\cal E}}
\newcommand{\cC}{{\cal C}}
\newcommand{\Ubs}{U_{\rm BS}}
\newcommand{\qq}{{\bf ???}}
\newcommand*{\etal}{\textit{et al.}}

\newcommand{\tcr}[1]{\textcolor{red}{#1}}
\newcommand{\tcb}[1]{\textcolor{blue}{#1}}
\newcommand{\tcm}[1]{\textcolor{magenta}{#1}}

\def\vec#1{\mathbf{#1}}
\def\ket#1{|#1\rangle}
\def\bra#1{\langle#1|}
\def\ketbra#1{|#1\rangle\langle#1|}
\def\braket#1{\langle#1|#1\rangle}
\newcommand{\scalp}[2]{\langle#1|#2\rangle}
\def\idmat{\mathbf{1}}
\def\caln{\mathcal{N}}
\def\calc{\mathcal{C}}
\def\rhon{\rho_{\mathcal{N}}}
\def\rhoc{\rho_{\mathcal{C}}}
\def\bfu{\mathbf{u}}
\def\bfmu{\mbox{\boldmath$\mu$}}

\newcommand{\be}{\begin{equation}}
\newcommand{\ee}{\end{equation}}
\newcommand{\bfg}{\begin{figure}}
\newcommand{\efg}{\end{figure}}
\newcommand{\Itwo}{\mathbb{1}_2}
\newcommand{\I}{\mathcal{I}}
\newcommand{\al}{\alpha}

\newcommand{\oost}{\frac{1}{\sqrt{2}}}
\newcommand{\rme}{\ensuremath{\mathrm{e}}}
\newcommand{\rmi}{\ensuremath{\mathrm{i}}}
\newcommand{\rmd}{\ensuremath{\mathrm{d}}}
\newcommand{\eg}{\textit{e.\thinspace g.\ }}
\newcommand{\ie}{\textit{i.\thinspace e.\ }}
\newcommand{\tr}{\operatorname{tr}}

\sloppy

\title{A heat bath can generate all classes of three-qubit entanglement
      } 
\author{Christopher Eltschka $^{1}$, Daniel Braun $^{2,3,4}$, and Jens Siewert $^{5,6}$}
\affiliation{$^1$ Institut f\"ur Theoretische Physik, 
         Universit\"at Regensburg, D-93040 Regensburg, Germany}
\affiliation{$^2$ Universit\'e Paul Sabatier, Toulouse, France}
\affiliation{$^3$ Laboratoire de Physique Th\'eorique, IRSAMC, UMR 5152 du CNRS} 
\affiliation{$^4$ Institut f\"ur theoretische Physik, Universit\"at T\"ubingen, D-72076 T\"ubingen, Germany}
\affiliation{$^5$ Departamento de Qu\'{\i}mica F\'{\i}sica, Universidad del Pa\'{\i}s Vasco 
            UPV/EHU, 48080 Bilbao, Spain}
\affiliation{$^6$ IKERBASQUE, Basque Foundation for Science, 48011 Bilbao, Spain}

\begin{abstract}
It is common knowledge that coupling to a heat bath, in general, 
tends to reduce the entanglement in a quantum system. In recent years,
increasing interest has been devoted to the opposite situation where 
thermal or specifically tailored environments may stabilize 
or even generate entanglement. We prove the universality
of this phenomenon for multipartite entanglement in 
the frame of an exactly solvable dephasing model.
We show by evaluating analytical lower bounds for the
appropriate entanglement measures that a common thermal environment 
can dynamically generate all entanglement classes 
of three qubits for almost any initial product state.
For sufficiently weak dissipation this entanglement
     may persist up to arbitrarily large times.
\end{abstract}
%
%
\maketitle

{\em Introduction.}---Entanglement, one of the hallmarks 
of quantum mechanics with a wide range
of applications in quantum information processing~\cite{Horodecki09}, 
is commonly thought of being rather fragile when 
exposed to decohering environments.
Yet, already towards the end of last century it turned out that
dissipative dynamics can lead to entanglement due to relaxation rates
that depend on the multipartite state~\cite{Plenio99},
stabilize entanglement in a decoherence-free subspace \cite{Beige00b}, or enable its
generation through local external driving~\cite{Beige00,schneider_entanglement_2002}. 
Later it was shown theoretically that specific  environments can be 
used to engineer  highly entangled states~\cite{Kraus2008,Verstraete2009},
with very recent experimental implementations~\cite{Wineland2013,Devoret2013}.
Apart from specifically tailoring dissipation it is interesting to ask about the action of 
a more generic or even thermal environment.
The somewhat surprising answer is  that the interaction with a common heat bath
{\em alone} can, 
under certain circumstances, entangle two subsystems~\cite{Braun02}. This kind of
``reservoir-induced entanglement'' has meanwhile been found for a large
variety of Markovian and non-Markovian evolutions, systems with finite
or infinite dimensional Hilbert spaces, many different environments (spins,
bosons, squeezed or thermal, single and many-mode), Brownian motion, 
and even vacuum or the Unruh radiation experienced by accelerated 
subsystems~\cite{kim_entanglement_2002,benatti_environment_2003,benatti_entanglement_2004,Benatti05,benatti_environment_2008,paz_dynamics_2008,ManiPlast2008,ManiPlast2009}.

The physical picture behind this effect is that the common heat bath not only 
leads to decoherence, but also to an effective interaction.  
Depending on the balance between the two, entanglement may arise, 
disappear again, or even persist in a steady state. 
With very few exceptions~\cite{Kraus2008,Verstraete2009,huai-xin_dynamics_2007,An_common_env_2007,li_multi-mode_2010,benatti_three_2011,pumulo_non-equilibrium_2011} 
these investigations were all concerned with bipartite entanglement.
However, entanglement in systems with more than two parties exhibits
a rich structure that is little explored. Three qubits represent 
the only multipartite system whose state space partitioning in terms of entanglement 
classes is completely known~\cite{Duer2000,Acin2001}. Thus the question arises
to what extent reservoir-induced entanglement generation is a universal phenomenon
and whether or not all entanglement types in a multipartite system can be generated
from separable states. 
The latest developments in entanglement theory have made it 
     feasible to give a quantitative answer.


Here we examine the generation of genuine tripartite entanglement for three qubits in a large
class of thermal dephasing environments using state of the art methods to detect 
the various entanglement resources.  We show that all entanglement classes can arise
from the coupling to the common heat bath alone and find the corresponding conditions for
the model parameters and initial states. We start our discussion by introducing the
dephasing models under consideration and solve the equations of motions exactly.
Before discussing in detail the entanglement classes 
in the system evolution we briefly review the entanglement
classification for three qubits and the appropriate measures 
to quantify these resources. 

{\em Model.}---Consider a system of qubits $H_{\text{sys}}$ 
interacting with a heat bath.  We will restrict ourselves to 
non-interacting
qubits with degenerate energy levels, i.e., $H_{\text{sys}}=0$,
so that the total Hamiltonian reads
\begin{equation} 
\label{H}
H=H_{\text{int}}+H_{\text{bath}}\ \ .
\end{equation}
This type of  model is also obtained from a more general dephasing
model (i.e., a model with $[H_{\text{sys}},H_{\text{int}}]=0$) by going to 
the interaction picture with respect to $H_{\text{sys}}$.
The system is coupled through an interaction Hamiltonian 
%
             $H_{\text{int}}=S\otimes B$
%
to a {\em common} heat bath, where the ``system coupling agent'' $S$  
acts on the qubit Hilbert space, and
``bath coupling agent'' $B$ on the bath degree of freedoms.
For the heat bath we assume a set of harmonic oscillators,
\begin{equation} 
\label{Hharm}
H_{\text{bath}}=\sum_j \left(\frac{p_j^2}{2m}+\frac{1}{2}m\omega_j^2q_j^2\right)
\end{equation}
and 
$B=\sum_j g_j q_j$
where $g_j$ are coupling constants to the $j$th oscillator.  
It is most
convenient to solve the dynamics of the resulting reduced density matrix in
the pointer basis $\{|s\rangle\}$ with $S|s\rangle=s|s\rangle$. With the
usual assumptions of factorizing initial conditions between system and bath,
and the bath in thermal equilibrium at temperature $T=1/k_B\beta$ at time $t=0$,
one finds the exact general 
solution in terms of the reduced density matrix $\rho(t)$ of the von
Neumann equation of motion   \cite{Braun01}, 
\begin{equation} \label{eq:vN}
\langle s|\rho(t)|s'\rangle=\re^{-(s-s')^2f(t)+\ri
(s^2-s'^2)\varphi(t)}\langle s|\rho(0)|s'\rangle\,, 
\end{equation}
where the functions $f(t)$ and $\varphi(t)$ are, respectively, related to
real and imaginary part of the thermal-bath correlation function:
\begin{eqnarray}
f(t) & = &          \sum_j\frac{g_j^2(1+2\overline{n}_j)}
              {2m\hbar\omega_j^3}(1-\cos\omega_j t) 
           =  \Re\int_0^t \frac{\text{d}s s}{\hbar^2} C(t-s)\nonumber\\
\varphi(t) & = &    \sum_j\frac{g_j^2(1+2\overline{n}_j)}
              {2m\hbar\omega_j^2}\left[t-\frac{\sin\omega_j t}{\omega_j}
                                 \right] 
           =  \Im\int_0^t   \frac{\text{d}s s}{\hbar^2} C(t-s)\nonumber\,,
\end{eqnarray}
$C(t)=\langle B(t)B(0)\rangle$ is evaluated in the initial thermal state of
  the heat bath, and $\overline{n}_j=1/(\re^{\beta \hbar 
  \omega_j}-1)$. 
Both $f(t)$ and $\varphi(t)$ vanish at $t=0$. For $t>0$ they 
are real, positive. 
While $\varphi(t)\propto t$ for large times, the behavior of  
$f(t)$ for large times depends on the heat bath.  If there is an ultraviolet 
cutoff,
$f(t)$ may saturate at some finite value, whereas otherwise $f(t)$ may 
diverge. 
Thus, the time evolution
arising from the action of a given heat bath
is characterized by a certain path $(f(t),\varphi(t))$ in the 
quadrant $\mathbb{R}_{\geq
0}\times\mathbb{R}_{\geq 0}$.

{\em Three qubits coupled to heat bath.}---%
In the appropriate local 
bases the most general form of $S$ reads
\begin{equation}
  \label{eq:generalS}
  S\ =\ \ri 
        \lambda_0\ \sigma_0^{(1)}\otimes \sigma_0^{(2)}\otimes \sigma_0^{(3)}
    \  +\ \sum_j \lambda_j \sigma_z^{(j)}
\end{equation}
%
where $\sigma_{\alpha}^{(j)}$ denotes the Pauli matrices 
for the $j$th qubit ($\alpha\in \{x,y,z\}$, $j\in\{1,2,3\}$),
$\sigma_0^{(j)}\equiv \ri\mathds{1}$,
and $\lambda_j\geq 0$.  We assume $\lambda_0=0$ because a finite
$\lambda_0$ amounts to shifting the bath oscillators and does not 
lead to qualitatively new behavior. Also, because
of the symmetry of the problem, we may 
assume $\lambda_1 \geq \lambda_2 \geq \lambda_3$.
Finally, a common multiplier of the $\lambda_j$ can be absorbed in the 
coupling constants $g_i$ and thus
just corresponds to  rescaling of
$f$ and $\varphi$, therefore we may choose $\lambda_1 = 1$. Thus the general
form 
of $S$ we use is
\begin{equation}
  \label{eq:S}
  S\ =\ \sigma_z^{(1)} + \lambda_2\sigma_z^{(2)} + \lambda_3\sigma_z^{(3)}, \quad 1 \geq \lambda_2 \geq \lambda_1 \geq 0
\end{equation}
with eigenvalues $s_{jkl}$
\begin{eqnarray}
\label{eq:evalsS}
   S\ket{jkl} & = & s_{jkl}\ \ket{jkl} 
\nonumber\\
              & = & \left((-1)^j+(-1)^k\lambda_2+(-1)^l\lambda_3\right)\ket{jkl}
\end{eqnarray}
which, substituted in Eq.~\eqref{eq:vN}, determine the time evolution
of the reduced system state.

{\em Quantifying entanglement resources.}---For three qubits, 
there are three types (or classes) 
of entanglement: Greenberger-Horne-Zeilinger (GHZ) type, $W$
type and biseparable (B)~\cite{Duer2000,Acin2001}. A mixed state is of a given
entanglement class if it has at least one decomposition that contains only
states of this class, but it has no decomposition with states only
from `lower' classes.
To date, there are no 
practicable methods to identify with certainty the entanglement type
of arbitrary mixed states. However, since the entanglement classes
form a hierarchy $\text{GHZ}\supset W\supset \text{B}$, 
it is meaningful to determine the class a state {\em at least} belongs to.

To identify GHZ-type entanglement, we use the three-tangle, which is
non-zero exactly for GHZ-type entangled states~\cite{CKW2000,Viehmann2012}. 
For pure three-qubit states $\psi$ it is defined as (we drop the qubit
index of the Pauli matrices)
\begin{equation}
\nonumber
   \tau_3 = \sqrt{\left| \sum_{j=0,x,z} \!\!
 \bra{\psi^{\ast}} \sigma_j\otimes\sigma_y\otimes\sigma_y\ket{\psi}\!
 \bra{\psi^{\ast}} \sigma_j\otimes\sigma_y\otimes\sigma_y\ket{\psi}
                 \right|
                      }
\end{equation}
(here $\psi^{\ast}$ denotes the state with complex conjugate coefficients)
and for mixed states $\rho$ with pure-state decompositions 
$\rho=\sum p_j\ket{\psi_j}\!\bra{\psi_j}$ it is the 
convex roof~\cite{Uhlmann1998}
\begin{equation}
     \tau_3(\rho)\ =\ \min_{\mathrm{\tiny all\ decomp.}} \sum\ p_j \ \tau_3(\psi_j)
\ \ .
\label{eq:convex_roof}
\end{equation}
While there is no
known way to calculate the exact three-tangle of arbitrary mixed
states, a lower bound can, in principle, 
be calculated analytically~\cite{ES2012-ScR}. If this
lower bound does not vanish, we know for sure 
the state is GHZ-type entangled. 

As for three qubits there are only two classes of genuine multipartite
entanglement (GME) the appropriate quantifier to measure the `$W$-ness' 
of a state is the GME concurrence~\cite{Ma2011}. To this end
we need to consider the three bipartitions 
$\{1|23\}$, $\{2|13\}$, and $\{3|12\}$ of qubit 1, qubit 2, and qubit 3.
For pure
states the GME concurrence is the minimum linear
entropy among all possible bipartitions $\gamma_i=\{A_i|B_i\}$ of a state 
\begin{equation}
  C_{\text{GME}}(\psi)=\min_{\gamma_i}\sqrt{2\left[1-\tr
                                      \left(\rho_{A_{i}}^2\right)\right]
                                           }
\label{eq:GME}
\end{equation}
where $\rho_{A_{i}}$ is the reduced density matrix of 
party $A_{i}$ in the bipartition $\gamma_i$. 
For mixed states, $C_{\text{GME}}$ is defined as the convex roof,
in analogy with Eq.~\eqref{eq:convex_roof}. For the explicit 
calculations we use the lower bound described in Ref.~\cite{Ma2011}.
The procedure to calculate the bounds involves an 
optimization in which we maximize the three-tangle or the GME
concurrence over local unitary transformations.

To determine whether the state is entangled at all we use the
negativity~\cite{Vidal2002}. That is, we consider again the bipartitions
$\gamma_i$, but now each of them separately. The negativity 
for the bipartition $\gamma_i$ is 
%
$   \mathcal{N}_{\gamma_i}(\rho)\ =\ 
             \frac{1}{2}\left( \|\rho^{T_{A_i}}\|_1-1 \right)
$
%
where $\rho^{T_{A_i}}$ denotes the partial transpose with respect to party ${A_i}$
and $\|\bullet\|_1$ is the trace norm.
Also $\mathcal{N}(\rho)$ can be regarded as
a lower bound as it does not detect entangled states with a positive
partial transpose. However, it  can
be computed easily.
Note that the maximum value of $\mathcal{N}$ for a single-qubit partition
is $\frac12$. As the maxima of $\tau_3$ and $C_{\text{GME}}$ equal $1$
we plot $2\mathcal{N}$.

{\em Equivalence of the evolutions of pure product states. }
Because we are interested in possible entanglement generation 
we study the time evolution starting with a fully separable
pure state, i.e., $\ket{\psi}=\ket{\phi_1}\ket{\phi_2}\ket{\phi_3}$ 
with $\ket{\phi_j} = \alpha_j\ket{0} + \beta_j\ket{1}$. 
All
product states can be obtained from the state $\ket{+} =
\oost\left(\ket0+\ket1\right)$ by the diagonal transformation $F =
\otimes_{j=1}^3 F_j$ with $F_j=\operatorname{diag}(\alpha_j, \beta_j)$ which, except for the special
cases $\ket{\phi_j}=\ket0$ and $\ket{\phi_j}=\ket1$, is
invertible. 
This in turn means that all tripartite product
states that contain neither $\ket0$ nor $\ket1$ as a factor are related
to each other by diagonal GL$(2,\mathbb{C})^{\otimes3}$ transformations.

Since we consider diagonal couplings $S$ to the bath 
the time evolution commutes with all diagonal
transformations. 
Thus, for any given $f$ and $\varphi$, the state
$\rho(f,\varphi;\phi_1\phi_2\phi_3)$ is related to the state 
$\rho(f,\varphi;{+}{+}{+})$ by the 
same diagonal transformation as the state $\ket{\phi_1\phi_2\phi_3}$ to the
state $\ket{{+}{+}{+}}$. In particular, as long as the initial state does not
contain a factor $\ket0$ or $\ket1$, for fixed $f$ and $\varphi$ all states
$\rho(f,\varphi;\phi_1\phi_2\phi_3)$ can be transformed {\em into one another}
by a GL$(2,\mathbb{C})^{\otimes3}$
transformation, and thus  belong to the same entanglement class. 
Consequently, it suffices to look at $\ket{{+}{+}{+}}$ to
capture the behavior of (almost) all initial product states.

We may apply analogous reasoning to GHZ-type entanglement and exploit
the properties  of the three-tangle under 
local transformations~\cite{Viehmann2012}.
For arbitrary $\rho$ and arbitrary local 
transformations $G\in\text{GL}(2,\mathbb{C})^{\otimes 3}$ we have 
$\tau_3(G\rho G^\dagger) = \left|\det(G)\right|^2 \tau_3(\rho)$. 
For a diagonal transformation $G$  which turns 
$\ket{{+}{+}{+}}$ into the normalized state $\ket{\phi_1\phi_2\phi_3}$ we find
$\left|\det G\right|^2 =
4\left|\prod_{j=1}^3\alpha_j\beta_j\right|^2 \leq 1$. 
Hence the state
$\rho(f,\varphi;+\!+\!+)$ has the largest three-tangle among all the states with the
same $f$ and $\varphi$. For any other state, the three-tangle can be
obtained by multiplication with the appropriate determinant.

In the case where one of the factors of the initial
state is $\ket0$ or $\ket1$, 
the transformation $F$ is basically a projection, so that
the qubit does not evolve at all. 
Thus, the problem effectively
reduces to the two-qubit case (or the one-qubit case if
two of the factors are $\ket0$ or $\ket1$).
Therefore, in what follows we will assume that the initial
state is 
%
\[
    \rho(t=0) \equiv\rho_0= \pi_{+++}\  \ \longrightarrow\  \
    \bra{jkl}\rho_0\ket{mnq}=\frac18\ \ ,
\]
%
with the abbreviation 
$\ket{\psi}\!\bra{\psi}\equiv\pi_{\psi}$. 
%

{\em Entanglement evolution in the $f$-$\varphi$ plane.}---The evolution
starts at $t=0$ with the pure product state $\pi_{+++}$ 
and $f=\varphi=0$.
While a state can never reach a point with $f=0$ but
$\varphi\neq0$, it is still worthwhile considering this case as in principle 
arbitrarily small values of $f$ are possible. 
Moreover, as the states at $f=0$ are pure,
all entanglement measures (which are continuous) can be evaluated exactly.
The three-tangle for $f=0$ is 
\[
\tau_3(f=0,\varphi) = 
\frac12\sqrt{\left|c_1 c_2 c_3 + \mathrm{i} s_1 s_2 s_3 - (c_1+c_2+c_3) + 2
\right|}
\]
 where $c_j = \cos(8\lambda_j\varphi(t))$, $s_j = \sin(8\lambda_j\varphi(t))$,
$j=1,2,3$ and $\lambda_1\equiv 1$.
For small $\varphi$, the leading term is
$\tau_3(\psi(\varphi)) = 8\sqrt{2}\lambda_2\lambda_3\left|\varphi\right|^{3/2} +
O(\left|\varphi\right|^{7/2})$. This shows that three-tangle can be produced
whenever both $\lambda_2$ and $\lambda_3$ are non-zero, that is, whenever all
three qubits are coupled to the bath. 
{\em Only} if both $\lambda_2$ and $\lambda_3$ are rational 
the $\varphi$ dependence is periodic.
In that case there exists a sequence of values $\varphi_n$ with
vanishing initial three-tangle $\tau_3(f=0,\varphi_n)=0$.
For $\varphi_n$ there is no entanglement at all in the system (``zero
lines'' of entanglement for all $f$ values).

On increasing $f$ the off-diagonal elements 
decay according to Eq.~\eqref{eq:vN} so that the states become more
and more mixed and the entanglement measures decrease. For our method
we always expect three-tangle and GME concurrence to vanish at finite
$f$ values. This is because the fidelity of the initial GHZ-entangled
state decreases with growing $f$ while our method does not detect $\tau_3$
or $C_{\text{GME}}$ for GHZ fidelities of the optimized state $<\frac12$. 
Nonetheless this shows that if,
for a given heat bath, $f(t\to\infty)$ saturates at a small enough
value, both types of 
genuine multipartite entanglement may persist up to arbitrarily large times.
The dynamics of bipartite entanglement will be discussed below.
The typical behavior of the entanglement measures as functions of $f$
and $\varphi$ for generic coupling parameters 
resembles the one displayed in Fig.~1.
\begin{figure}[bh]
  \centering
  \includegraphics[width=.95\linewidth]{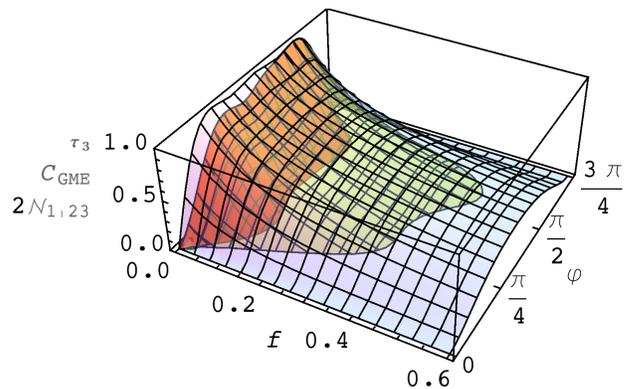}
  \caption{ Entanglement classes GHZ, $W$, B (red, yellow, light blue)
           for the case $\lambda_2=\frac23$, $\lambda_3=\frac13$. 
           Qualitatively the behavior resembles that of the generic
           case, only that here there is a periodicity in $\varphi$
           with zero lines for the entanglement measures
           at the values of $\varphi=3n\pi/4$.
           The negativity is calculated for the bipartition
           $\{1|23\}$, i.e., for the first qubit.
                     }
  \label{fig:classes1313}
\end{figure}

{\em The states for $f\to\infty$.}---%
It is useful to study the behavior for $f\to\infty$
as it reveals the special cases that need to be considered.
Recall that $f(t\to\infty)$  may diverge if not prevented by a cutoff.
We conclude from Eq.~\eqref{eq:vN} that
%
%
$\rho(f\to\infty) = \sum_rP_r\rho_0P_r^{\dagger}$ 
where $P_r$ are the
projectors on the eigenspaces of $S$. 
We have to determine
those eigenspaces depending on $\lambda_2$ and $\lambda_3$. 
Since $S$ is diagonal in the computational 
basis, $\ket{jkl}$ are trivially eigenstates, 
with corresponding eigenvalues $s_{jkl}$ according to Eq.~\eqref{eq:evalsS}.
As we start with the pure state
$\ket{{+}{+}{+}}$, the asymptotic state $\rho(f\to\infty)$ 
is a mixture of pure states that are equally-weighted
superpositions of all the basis states in the corresponding
eigenspace. Note that all entries of $\rho(f\to\infty)$ are either $\frac18$ or
$0$.

For generic values of $\lambda_2$ and $\lambda_3$,  all eigenspaces
are one-dimensional, and $\rho(f\to\infty)$ is the completely mixed state.
Since there is a finite neighborhood of the completely mixed state where all
states are separable \cite{BCJLPS1999}, 
generically full separability will be reached for finite values of $f$.
However, for certain values of $\lambda_2$ and $\lambda_3$, some of the
eigenvalues will coincide, resulting in larger eigenspaces and thus
more interesting asymptotic states.

In the trivial case $\lambda_2=\lambda_3=0$ no entanglement is generated. 
If only $\lambda_3=0$ and $\lambda_2>0$ the third
qubit is decoupled from the bath and remains in 
$\ket{+}$ while the first two qubits follow the well-known two-qubit
behavior~\cite{Braun02}: 
For generic $\lambda_2$ and  \mbox{$f\to\infty$} the state of the first two qubits 
is completely mixed and separable states are reached for finite $f$,
while in the symmetric case $\lambda_2=1$ 
the final state of the first two qubits is the separable mixture
\begin{equation}
  \label{eq:rho2s}
   \rho_{2s}\ =\ \frac12 \pi_{\psi^+}\ +\ \frac14\left(\pi_{00}+\pi_{11}\right)
\end{equation}
with $\ket{\psi^+}=\oost(\ket{01}+\ket{10})$.

Special cases arise for all $\lambda_j\neq 0$ 
if $\lambda_2=\lambda_3$,
$\lambda_2+\lambda_3=1$  or $\lambda_2=1$. In the first case, the eigenvalues for the
states $\ket{001}$ and $\ket{010}$ as well as those for $\ket{101}$ and
$\ket{110}$ coincide. For $\lambda_2+\lambda_3=1$, the states $\ket{011}$ and
$\ket{100}$ have the same eigenvalues. Finally, in the case $\lambda_2=1$
there is a degeneracy for the states $\ket{010}$ and $\ket{100}$ as
well as for the states $\ket{011}$ and $\ket{101}$.
Not all  of the three relations above can be fulfilled at the
same time. Moreover, $\lambda_2=1, \lambda_3\neq1$ is
equivalent up to rescaling to $\lambda_2=\lambda_3\neq1$ after exchanging
the first with the third qubit. Finally, if the second and third
equalities are fulfilled, we have $\lambda_3=0$ which decouples the third
qubit. Hence there remain 
four (truly tripartite) special cases with the corresponding
asymptotic states $\rho(f\to\infty)$, 
which are located at the border between separable and biseparable
                      states, see Table I.
The negativity is non-zero for $t\to\infty$ for at least one
bipartition in each of those special cases (cf.~Fig.~2).
\begin{table}[h]
\caption{\label{tab:1} The four special cases for $\rho(f\to\infty)$
                      depending on the parameters $\lambda_2,\lambda_3$,
                      and the generic case where $(\lambda_2,\lambda_3)$
                      differs from the values in the special cases.
                      The states used in the decompositions are
$\ket{W} =
\frac{1}{\sqrt3}\left(\ket{001}+\ket{010}+\ket{100}\right)$, 
$\ket{\overline W} = \frac{1}{\sqrt3}\left(\ket{110}+\ket{101}+\ket{011}\right)$, 
                      and
$\ket{\mathrm{GHZ}_3^+}=\oost(\ket{011}+\ket{100})$.
                      All asymptotic states are separable. 
        }
\begin{ruledtabular}
\begin{tabular}{l|c}
parameters  & asymptotic state $\rho(f\to\infty)$
\\[1mm] \hline\\
$\lambda_2 + \lambda_3 = 1$        & $\frac14 \pi_{\mathrm{GHZ}_3^+} 
             +\frac18  \left(\pi_{000}+\pi_{001}+\pi_{010}+
                    \right.$
\\ 
$\lambda_j \notin \{0,\frac12,1\}$ &
                                     $\left. \ \ \ \ \ \ \ \ \ \ 
                                             \ \ \ \ \ \ \ \ \ \ 
                                      +\pi_{101}+\pi_{110}+\pi_{111}
                                      \right)$
\\[1mm] \hline\\
$\lambda_2 = \lambda_3 = 1$   & $\frac18 \pi_{000} + \frac38 \pi_{W} 
                                                + \frac38 \pi_{\overline W} 
                                                + \frac18 \pi_{111}$
\\[1mm] \hline\\
$\lambda_2 = \lambda_3 = \frac12$  &
                     $\frac14 \left(\pi_{\mathrm{GHZ}_3^+} + \pi_{0\psi^+} +
                      \pi_{1\psi^+}\right) + 
                      \frac18 \left(\pi_{000} + \pi_{111}\right)$
\\[1mm] \hline\\
$\lambda_2 = \lambda_3$            & 
                            $\frac12(\pi_0 + \pi_1)\otimes\rho_{2s}$
\\
$\notin \{0,\frac12,1\}$           &
\\[1mm] \hline\\
otherwise
                                   & $\frac{1}{8}
                                     \mathds{1}\otimes\mathds{1}\otimes\mathds{1}$
\end{tabular}
\end{ruledtabular}
\end{table}
\begin{figure}
  \centering
  \includegraphics[width=.32\linewidth]{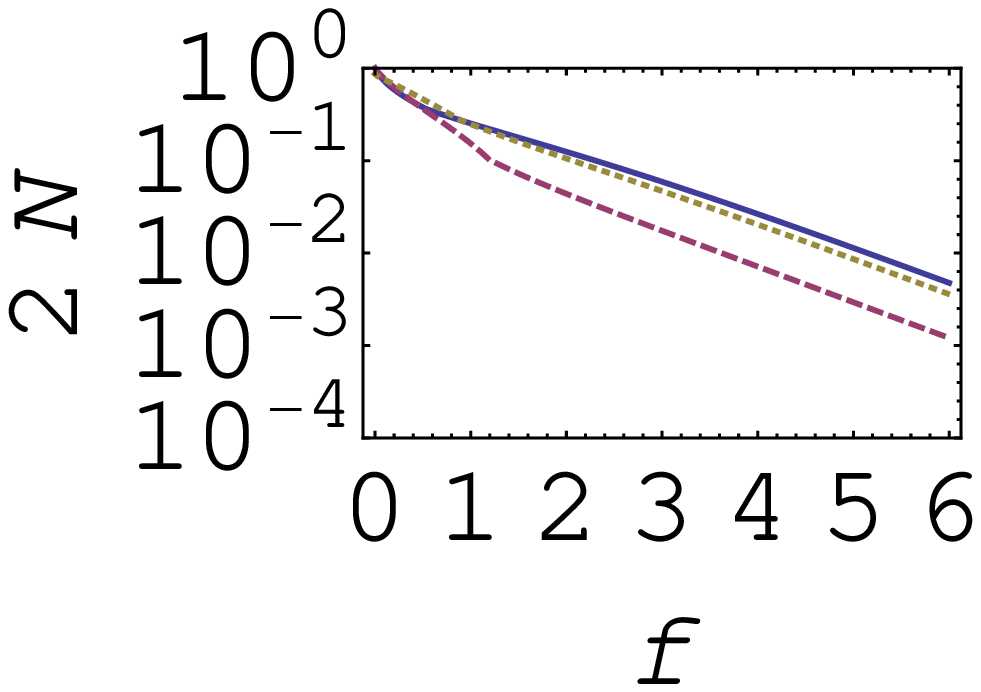}
  \includegraphics[width=.32\linewidth]{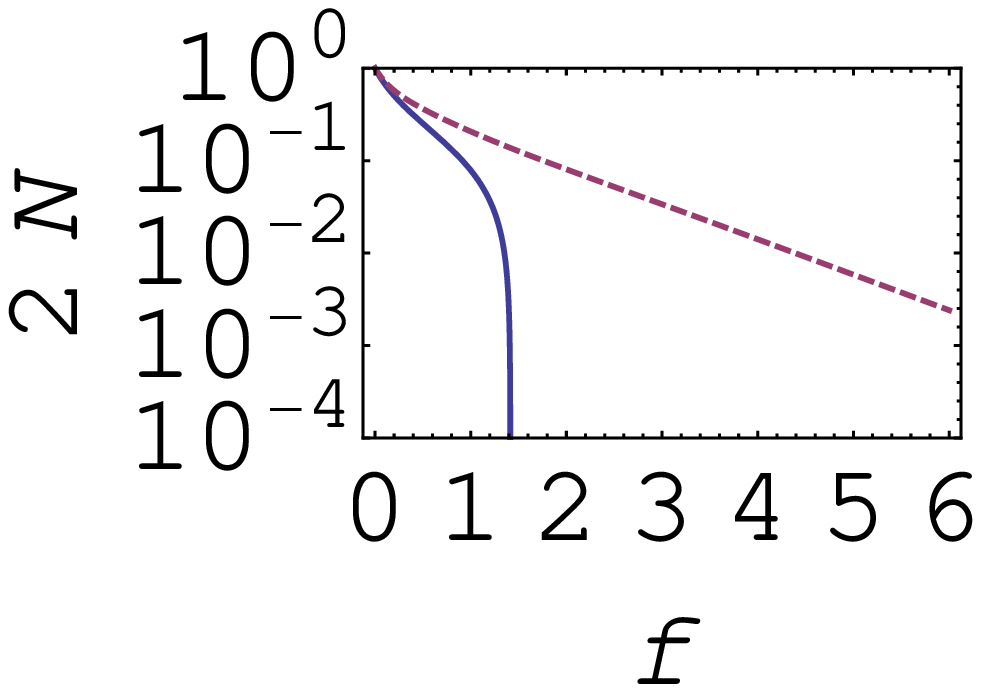}
  \includegraphics[width=.32\linewidth]{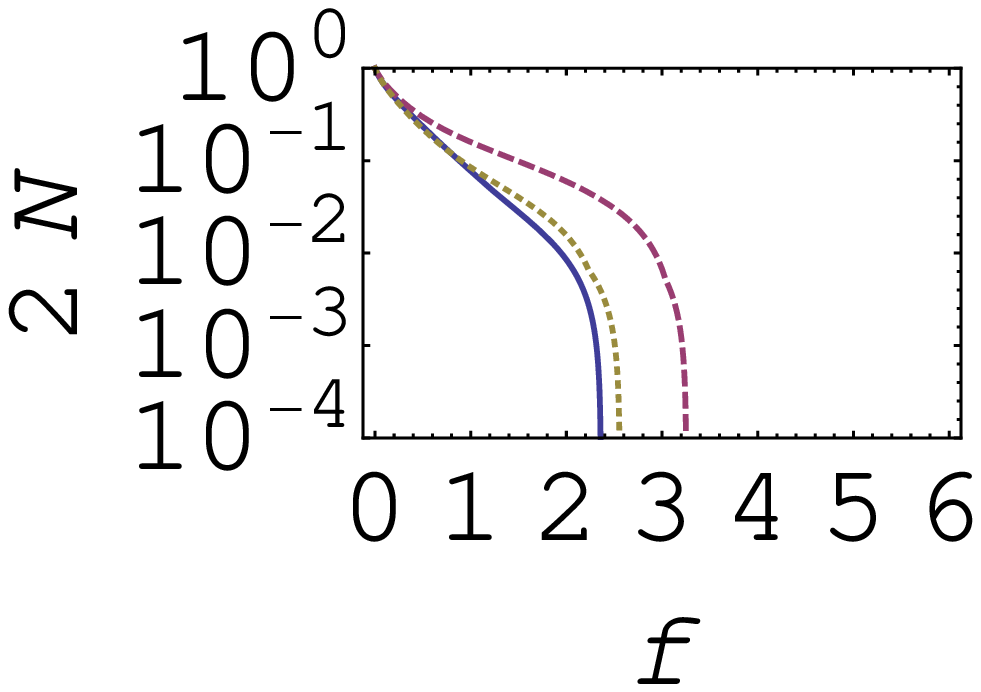}
  \caption{ Negativities $2\mathcal{N}(f)$ 
           in the asymptotic states
           according to Table~1 for values of $\varphi$ maximizing
           $\mathcal{N}(f=0)$. In each plot, the blue solid line gives the
           negativity for the bipartition $\{1|23\}$, the red long-dashed line
           for the bipartition $\{2|13\}$, and the green dashed line for the
           bipartition $\{3|12\}$.
\\
           a) For $\lambda_2+\lambda_3=1$ (first line of Table~1)
           the negativities decay exponentially, with a specific
           decay constant for each qubit. We do not show examples
           for the second and the third line of Table~1, since
           the only difference with the first line
           is that all qubits have
           the same decay constant ($\lambda_2=\lambda_3=1$), or
           only the decay constant of the first qubit differs
           from the others ($\lambda_2=\lambda_3=\frac{1}{2}$).
           The plot shows the negativities for $\lambda_2 = 2/3$,
             $\lambda_3 = 1/3$ and $\varphi = 3\pi/8$.
\\
           b) For $\lambda_2=\lambda_3\notin \{0,\frac12,1\}$
           the negativity of the first qubit vanishes at a finite
           value of $f$ while that of the other qubits
           vanishes smoothly.
           Plot parameters are
           $\lambda_2 =\lambda_3 = 1/3$ and $\varphi = 3\pi/8$.
\\
           c) In the generic case all negativities vanish at 
           finite values of $f$
           (here for $\lambda_2 =\pi/4$, 
           $\lambda_3 = \mathrm{e}/4$ and $\varphi = 1/2$).
    }
  \label{fig:neg11}
\end{figure}

{\em Summary.}---We have analyzed multipartite
entanglement generation in 
an exactly solvable dephasing model of three non-interacting qubits coupled
to the same  thermal heat bath. Entanglement generation
solely depends on the interplay between the real and imaginary parts $f(t)$ and
$\varphi(t)$ of the
bath-correlation function, and two parameters $\lambda_2$, $\lambda_3$ of the
coupling Hamiltonian. 

Using  optimized  lower
bounds to the three-tangle and the GME 
concurrence, we have shown that for small $f(t)$ (weak decoherence) and
sufficiently large $\varphi(t)$ (strong 
effective interaction mediated by the heat 
bath) all classes of tripartite entanglement can be generated from almost
any pure product state. For almost all pairs $(\lambda_2,\lambda_3)$ the
state for sufficiently large finite $f$ is  fully separable, but there
are four special symmetric cases, where the reservoir-induced 
bipartite entanglement
 is proven to persist up to arbitrarily large values of $f(t)$.
If the bath correlation function $f(t)$ for $t\to\infty$
saturates at sufficiently small values, genuine tripartite and even
GHZ-type entanglement may be present up to arbitrarily large times.
One may note that these statements hold as well for weakly
mixed initial separable states.

{\em Acknowledgements.}---This work was funded by the 
German Research Foundation within 
SPP 1386 (C.E.), by Basque Government grant IT-472-10 
and MINECO grant FIS2012-36673-C03-01 (J.S.).
C.E.\ and J.S.\ thank  J.\ Fabian and K.\ Richter for their support.

%
%
%
%

\bibliography{./mybibs_bt}

\begin{thebibliography}{33}
\expandafter\ifx\csname natexlab\endcsname\relax\def\natexlab#1{#1}\fi
\expandafter\ifx\csname bibnamefont\endcsname\relax
  \def\bibnamefont#1{#1}\fi
\expandafter\ifx\csname bibfnamefont\endcsname\relax
  \def\bibfnamefont#1{#1}\fi
\expandafter\ifx\csname citenamefont\endcsname\relax
  \def\citenamefont#1{#1}\fi
\expandafter\ifx\csname url\endcsname\relax
  \def\url#1{\texttt{#1}}\fi
\expandafter\ifx\csname urlprefix\endcsname\relax\def\urlprefix{URL }\fi
\providecommand{\bibinfo}[2]{#2}
\providecommand{\eprint}[2][]{\url{#2}}

\bibitem[{\citenamefont{Horodecki et~al.}(2009)\citenamefont{Horodecki,
  Horodecki, Horodecki, and Horodecki}}]{Horodecki09}
\bibinfo{author}{\bibfnamefont{R.}~\bibnamefont{Horodecki}},
  \bibinfo{author}{\bibfnamefont{P.}~\bibnamefont{Horodecki}},
  \bibinfo{author}{\bibfnamefont{M.}~\bibnamefont{Horodecki}},
  \bibnamefont{and}
  \bibinfo{author}{\bibfnamefont{K.}~\bibnamefont{Horodecki}},
  \bibinfo{journal}{Rev. Mod. Phys.} \textbf{\bibinfo{volume}{81}},
  \bibinfo{pages}{865} (\bibinfo{year}{2009}).

\bibitem[{\citenamefont{Plenio et~al.}(1999)\citenamefont{Plenio, Huelga,
  Beige, and Knight}}]{Plenio99}
\bibinfo{author}{\bibfnamefont{M.~B.} \bibnamefont{Plenio}},
  \bibinfo{author}{\bibfnamefont{S.~F.} \bibnamefont{Huelga}},
  \bibinfo{author}{\bibfnamefont{A.}~\bibnamefont{Beige}}, \bibnamefont{and}
  \bibinfo{author}{\bibfnamefont{P.~L.} \bibnamefont{Knight}},
  \bibinfo{journal}{Phys. Rev. A} \textbf{\bibinfo{volume}{59}},
  \bibinfo{pages}{2468} (\bibinfo{year}{1999}).

\bibitem[{\citenamefont{Beige et~al.}(2000{\natexlab{a}})\citenamefont{Beige,
  Braun, and Knight}}]{Beige00b}
\bibinfo{author}{\bibfnamefont{A.}~\bibnamefont{Beige}},
  \bibinfo{author}{\bibfnamefont{D.}~\bibnamefont{Braun}}, \bibnamefont{and}
  \bibinfo{author}{\bibfnamefont{P.~L.} \bibnamefont{Knight}},
  \bibinfo{journal}{New Journal of Physics} \textbf{\bibinfo{volume}{2}},
  \bibinfo{pages}{22.1} (\bibinfo{year}{2000}{\natexlab{a}}).

\bibitem[{\citenamefont{Beige et~al.}(2000{\natexlab{b}})\citenamefont{Beige,
  Braun, Tregenna, and Knight}}]{Beige00}
\bibinfo{author}{\bibfnamefont{A.}~\bibnamefont{Beige}},
  \bibinfo{author}{\bibfnamefont{D.}~\bibnamefont{Braun}},
  \bibinfo{author}{\bibfnamefont{B.}~\bibnamefont{Tregenna}}, \bibnamefont{and}
  \bibinfo{author}{\bibfnamefont{P.~L.} \bibnamefont{Knight}},
  \bibinfo{journal}{Phys. Rev. Lett.} \textbf{\bibinfo{volume}{85}},
  \bibinfo{pages}{1762} (\bibinfo{year}{2000}{\natexlab{b}}).

\bibitem[{\citenamefont{Schneider and
  Milburn}(2002)}]{schneider_entanglement_2002}
\bibinfo{author}{\bibfnamefont{S.}~\bibnamefont{Schneider}} \bibnamefont{and}
  \bibinfo{author}{\bibfnamefont{G.~J.} \bibnamefont{Milburn}},
  \bibinfo{journal}{Phys. Rev. A} \textbf{\bibinfo{volume}{65}}
  (\bibinfo{year}{2002}).

\bibitem[{\citenamefont{Kraus et~al.}(2008)\citenamefont{Kraus, B{\"u}chler,
  Diehl, Kantian, Micheli, and Zoller}}]{Kraus2008}
\bibinfo{author}{\bibfnamefont{B.}~\bibnamefont{Kraus}},
  \bibinfo{author}{\bibfnamefont{H.~P.} \bibnamefont{B{\"u}chler}},
  \bibinfo{author}{\bibfnamefont{S.}~\bibnamefont{Diehl}},
  \bibinfo{author}{\bibfnamefont{A.}~\bibnamefont{Kantian}},
  \bibinfo{author}{\bibfnamefont{A.}~\bibnamefont{Micheli}}, \bibnamefont{and}
  \bibinfo{author}{\bibfnamefont{P.}~\bibnamefont{Zoller}},
  \bibinfo{journal}{\pra} \textbf{\bibinfo{volume}{78}},
  \bibinfo{pages}{042307} (\bibinfo{year}{2008}).

\bibitem[{\citenamefont{Verstraete et~al.}(2009)\citenamefont{Verstraete, Wolf,
  and Cirac}}]{Verstraete2009}
\bibinfo{author}{\bibfnamefont{F.}~\bibnamefont{Verstraete}},
  \bibinfo{author}{\bibfnamefont{M.~M.} \bibnamefont{Wolf}}, \bibnamefont{and}
  \bibinfo{author}{\bibfnamefont{J.~I.} \bibnamefont{Cirac}},
  \bibinfo{journal}{Nature Physics}  (\bibinfo{year}{2009}).

\bibitem[{\citenamefont{Lin et~al.}(2013)\citenamefont{Lin, Gaebler, Reiter,
  Tan, Bowler, S{\o}rensen, Leibfried, and Wineland}}]{Wineland2013}
\bibinfo{author}{\bibfnamefont{Y.}~\bibnamefont{Lin}},
  \bibinfo{author}{\bibfnamefont{J.~P.} \bibnamefont{Gaebler}},
  \bibinfo{author}{\bibfnamefont{F.}~\bibnamefont{Reiter}},
  \bibinfo{author}{\bibfnamefont{T.~R.} \bibnamefont{Tan}},
  \bibinfo{author}{\bibfnamefont{R.}~\bibnamefont{Bowler}},
  \bibinfo{author}{\bibfnamefont{A.~S.} \bibnamefont{S{\o}rensen}},
  \bibinfo{author}{\bibfnamefont{D.}~\bibnamefont{Leibfried}},
  \bibnamefont{and} \bibinfo{author}{\bibfnamefont{D.~J.}
  \bibnamefont{Wineland}}, \bibinfo{journal}{Nature}
  \textbf{\bibinfo{volume}{504}}, \bibinfo{pages}{415} (\bibinfo{year}{2013}).

\bibitem[{\citenamefont{Shankar et~al.}(2013)\citenamefont{Shankar, Hatridge,
  Leghtas, Sliwa, Narla, Vool, Girvin, Frunzio, Mirrahimi, and
  Devoret}}]{Devoret2013}
\bibinfo{author}{\bibfnamefont{S.}~\bibnamefont{Shankar}},
  \bibinfo{author}{\bibfnamefont{M.}~\bibnamefont{Hatridge}},
  \bibinfo{author}{\bibfnamefont{Z.}~\bibnamefont{Leghtas}},
  \bibinfo{author}{\bibfnamefont{K.~M.} \bibnamefont{Sliwa}},
  \bibinfo{author}{\bibfnamefont{A.}~\bibnamefont{Narla}},
  \bibinfo{author}{\bibfnamefont{U.}~\bibnamefont{Vool}},
  \bibinfo{author}{\bibfnamefont{S.~M.} \bibnamefont{Girvin}},
  \bibinfo{author}{\bibfnamefont{L.}~\bibnamefont{Frunzio}},
  \bibinfo{author}{\bibfnamefont{M.}~\bibnamefont{Mirrahimi}},
  \bibnamefont{and} \bibinfo{author}{\bibfnamefont{M.~H.}
  \bibnamefont{Devoret}}, \bibinfo{journal}{Nature}
  \textbf{\bibinfo{volume}{504}}, \bibinfo{pages}{419} (\bibinfo{year}{2013}).

\bibitem[{\citenamefont{Braun}(2002)}]{Braun02}
\bibinfo{author}{\bibfnamefont{D.}~\bibnamefont{Braun}},
  \bibinfo{journal}{Phys. Rev. Lett.} \textbf{\bibinfo{volume}{89}},
  \bibinfo{pages}{277901} (\bibinfo{year}{2002}).

\bibitem[{\citenamefont{Kim et~al.}(2002)\citenamefont{Kim, Lee, Ahn, and
  Knight}}]{kim_entanglement_2002}
\bibinfo{author}{\bibfnamefont{M.~S.} \bibnamefont{Kim}},
  \bibinfo{author}{\bibfnamefont{J.}~\bibnamefont{Lee}},
  \bibinfo{author}{\bibfnamefont{D.}~\bibnamefont{Ahn}}, \bibnamefont{and}
  \bibinfo{author}{\bibfnamefont{P.~L.} \bibnamefont{Knight}},
  \bibinfo{journal}{Phys. Rev. A} \textbf{\bibinfo{volume}{65}}
  (\bibinfo{year}{2002}).

\bibitem[{\citenamefont{Benatti et~al.}(2003)\citenamefont{Benatti, Floreanini,
  and Piani}}]{benatti_environment_2003}
\bibinfo{author}{\bibfnamefont{F.}~\bibnamefont{Benatti}},
  \bibinfo{author}{\bibfnamefont{R.}~\bibnamefont{Floreanini}},
  \bibnamefont{and} \bibinfo{author}{\bibfnamefont{M.}~\bibnamefont{Piani}},
  \bibinfo{journal}{Phys. Rev. Lett.} \textbf{\bibinfo{volume}{91}},
  \bibinfo{pages}{070402} (\bibinfo{year}{2003}).

\bibitem[{\citenamefont{Benatti and
  Floreanini}(2004)}]{benatti_entanglement_2004}
\bibinfo{author}{\bibfnamefont{F.}~\bibnamefont{Benatti}} \bibnamefont{and}
  \bibinfo{author}{\bibfnamefont{R.}~\bibnamefont{Floreanini}},
  \bibinfo{journal}{Phys. Rev. A} \textbf{\bibinfo{volume}{70}},
  \bibinfo{pages}{012112} (\bibinfo{year}{2004}).

\bibitem[{\citenamefont{Benatti and Floreanini}(2005)}]{Benatti05}
\bibinfo{author}{\bibfnamefont{F.}~\bibnamefont{Benatti}} \bibnamefont{and}
  \bibinfo{author}{\bibfnamefont{R.}~\bibnamefont{Floreanini}},
  \bibinfo{journal}{Int. J. Mod. Phys. B} \textbf{\bibinfo{volume}{19}},
  \bibinfo{pages}{3063} (\bibinfo{year}{2005}).

\bibitem[{\citenamefont{Benatti et~al.}(2008)\citenamefont{Benatti, Liguori,
  and Nagy}}]{benatti_environment_2008}
\bibinfo{author}{\bibfnamefont{F.}~\bibnamefont{Benatti}},
  \bibinfo{author}{\bibfnamefont{A.~M.} \bibnamefont{Liguori}},
  \bibnamefont{and} \bibinfo{author}{\bibfnamefont{A.}~\bibnamefont{Nagy}},
  \bibinfo{journal}{J. Math. Phys.} \textbf{\bibinfo{volume}{49}},
  \bibinfo{pages}{042103} (\bibinfo{year}{2008}).

\bibitem[{\citenamefont{Paz and Roncaglia}(2008)}]{paz_dynamics_2008}
\bibinfo{author}{\bibfnamefont{J.~P.} \bibnamefont{Paz}} \bibnamefont{and}
  \bibinfo{author}{\bibfnamefont{A.~J.} \bibnamefont{Roncaglia}},
  \bibinfo{journal}{Phys. Rev. Lett.} \textbf{\bibinfo{volume}{100}},
  \bibinfo{pages}{220401} (\bibinfo{year}{2008}).

\bibitem[{\citenamefont{Maniscalco et~al.}(2008)\citenamefont{Maniscalco,
  Francica, Zaffino, Gullo, and Plastina}}]{ManiPlast2008}
\bibinfo{author}{\bibfnamefont{S.}~\bibnamefont{Maniscalco}},
  \bibinfo{author}{\bibfnamefont{F.}~\bibnamefont{Francica}},
  \bibinfo{author}{\bibfnamefont{R.~L.} \bibnamefont{Zaffino}},
  \bibinfo{author}{\bibfnamefont{N.~L.} \bibnamefont{Gullo}}, \bibnamefont{and}
  \bibinfo{author}{\bibfnamefont{F.}~\bibnamefont{Plastina}},
  \bibinfo{journal}{Phys.\ Rev.\ Lett.} \textbf{\bibinfo{volume}{100}},
  \bibinfo{pages}{090503} (\bibinfo{year}{2008}).

\bibitem[{\citenamefont{Francica et~al.}(2009)\citenamefont{Francica,
  Maniscalco, Piilo, Plastina, and Suominen}}]{ManiPlast2009}
\bibinfo{author}{\bibfnamefont{F.}~\bibnamefont{Francica}},
  \bibinfo{author}{\bibfnamefont{S.}~\bibnamefont{Maniscalco}},
  \bibinfo{author}{\bibfnamefont{J.}~\bibnamefont{Piilo}},
  \bibinfo{author}{\bibfnamefont{F.}~\bibnamefont{Plastina}}, \bibnamefont{and}
  \bibinfo{author}{\bibfnamefont{K.-A.} \bibnamefont{Suominen}},
  \bibinfo{journal}{Phys.\ Rev.\ A} \textbf{\bibinfo{volume}{79}},
  \bibinfo{pages}{032310} (\bibinfo{year}{2009}).

\bibitem[{\citenamefont{Huai-Xin}(2007)}]{huai-xin_dynamics_2007}
\bibinfo{author}{\bibfnamefont{L.}~\bibnamefont{Huai-Xin}},
  \bibinfo{journal}{Chinese Phys.} \textbf{\bibinfo{volume}{16}},
  \bibinfo{pages}{1878} (\bibinfo{year}{2007}).

\bibitem[{\citenamefont{An et~al.}(2007)\citenamefont{An, Wang, and
  Luo}}]{An_common_env_2007}
\bibinfo{author}{\bibfnamefont{J.-H.} \bibnamefont{An}},
  \bibinfo{author}{\bibfnamefont{S.-J.} \bibnamefont{Wang}}, \bibnamefont{and}
  \bibinfo{author}{\bibfnamefont{H.-G.} \bibnamefont{Luo}},
  \bibinfo{journal}{Physica A} \textbf{\bibinfo{volume}{382}}
  (\bibinfo{year}{2007}).

\bibitem[{\citenamefont{Li et~al.}(2010)\citenamefont{Li, Sun, and
  Ficek}}]{li_multi-mode_2010}
\bibinfo{author}{\bibfnamefont{G.-X.} \bibnamefont{Li}},
  \bibinfo{author}{\bibfnamefont{L.-H.} \bibnamefont{Sun}}, \bibnamefont{and}
  \bibinfo{author}{\bibfnamefont{Z.}~\bibnamefont{Ficek}}, \bibinfo{journal}{J.
  Phys. B: At. Mol. Opt. Phys.} \textbf{\bibinfo{volume}{43}},
  \bibinfo{pages}{135501} (\bibinfo{year}{2010}).

\bibitem[{\citenamefont{Benatti and Nagy}(2011)}]{benatti_three_2011}
\bibinfo{author}{\bibfnamefont{F.}~\bibnamefont{Benatti}} \bibnamefont{and}
  \bibinfo{author}{\bibfnamefont{A.}~\bibnamefont{Nagy}},
  \bibinfo{journal}{Annals of Physics} \textbf{\bibinfo{volume}{326}},
  \bibinfo{pages}{740} (\bibinfo{year}{2011}).

\bibitem[{\citenamefont{Pumulo et~al.}(2011)\citenamefont{Pumulo, Sinayskiy,
  and Petruccione}}]{pumulo_non-equilibrium_2011}
\bibinfo{author}{\bibfnamefont{N.}~\bibnamefont{Pumulo}},
  \bibinfo{author}{\bibfnamefont{I.}~\bibnamefont{Sinayskiy}},
  \bibnamefont{and}
  \bibinfo{author}{\bibfnamefont{F.}~\bibnamefont{Petruccione}},
  \bibinfo{journal}{Physics Letters A} \textbf{\bibinfo{volume}{375}},
  \bibinfo{pages}{3157} (\bibinfo{year}{2011}).

\bibitem[{\citenamefont{D\"ur et~al.}(2000)\citenamefont{D\"ur, Vidal, and
  Cirac}}]{Duer2000}
\bibinfo{author}{\bibfnamefont{W.}~\bibnamefont{D\"ur}},
  \bibinfo{author}{\bibfnamefont{G.}~\bibnamefont{Vidal}}, \bibnamefont{and}
  \bibinfo{author}{\bibfnamefont{J.~I.} \bibnamefont{Cirac}},
  \bibinfo{journal}{Phys.\ Rev.\ A} \textbf{\bibinfo{volume}{62}},
  \bibinfo{pages}{062314} (\bibinfo{year}{2000}).

\bibitem[{\citenamefont{Acin et~al.}(2001)\citenamefont{Acin, Bru{\ss},
  Lewenstein, and Sanpera}}]{Acin2001}
\bibinfo{author}{\bibfnamefont{A.}~\bibnamefont{Acin}},
  \bibinfo{author}{\bibfnamefont{D.}~\bibnamefont{Bru{\ss}}},
  \bibinfo{author}{\bibfnamefont{M.}~\bibnamefont{Lewenstein}},
  \bibnamefont{and} \bibinfo{author}{\bibfnamefont{A.}~\bibnamefont{Sanpera}},
  \bibinfo{journal}{Phys.\ Rev.\ Lett.} \textbf{\bibinfo{volume}{87}},
  \bibinfo{pages}{040401} (\bibinfo{year}{2001}).

\bibitem[{\citenamefont{Braun et~al.}(2001)\citenamefont{Braun, Haake, and
  Strunz}}]{Braun01}
\bibinfo{author}{\bibfnamefont{D.}~\bibnamefont{Braun}},
  \bibinfo{author}{\bibfnamefont{F.}~\bibnamefont{Haake}}, \bibnamefont{and}
  \bibinfo{author}{\bibfnamefont{W.}~\bibnamefont{Strunz}},
  \bibinfo{journal}{Phys. Rev. Lett.} \textbf{\bibinfo{volume}{86}},
  \bibinfo{pages}{2913} (\bibinfo{year}{2001}).

\bibitem[{\citenamefont{Coffman et~al.}(2000)\citenamefont{Coffman, Kundu, and
  Wootters}}]{CKW2000}
\bibinfo{author}{\bibfnamefont{V.}~\bibnamefont{Coffman}},
  \bibinfo{author}{\bibfnamefont{J.}~\bibnamefont{Kundu}}, \bibnamefont{and}
  \bibinfo{author}{\bibfnamefont{W.}~\bibnamefont{Wootters}},
  \bibinfo{journal}{Phys.\ Rev.\ A} \textbf{\bibinfo{volume}{61}},
  \bibinfo{pages}{052306} (\bibinfo{year}{2000}).

\bibitem[{\citenamefont{Viehmann et~al.}(2012)\citenamefont{Viehmann, Eltschka,
  and Siewert}}]{Viehmann2012}
\bibinfo{author}{\bibfnamefont{O.}~\bibnamefont{Viehmann}},
  \bibinfo{author}{\bibfnamefont{C.}~\bibnamefont{Eltschka}}, \bibnamefont{and}
  \bibinfo{author}{\bibfnamefont{J.}~\bibnamefont{Siewert}},
  \bibinfo{journal}{Appl.\ Phys.\ B} \textbf{\bibinfo{volume}{106}},
  \bibinfo{pages}{533} (\bibinfo{year}{2012}).

\bibitem[{\citenamefont{Uhlmann}(1998)}]{Uhlmann1998}
\bibinfo{author}{\bibfnamefont{A.}~\bibnamefont{Uhlmann}},
  \bibinfo{journal}{Open Sys.\ \& Inf.\ Dyn.} \textbf{\bibinfo{volume}{5}},
  \bibinfo{pages}{209} (\bibinfo{year}{1998}).

\bibitem[{\citenamefont{Eltschka and Siewert}(2012)}]{ES2012-ScR}
\bibinfo{author}{\bibfnamefont{C.}~\bibnamefont{Eltschka}} \bibnamefont{and}
  \bibinfo{author}{\bibfnamefont{J.}~\bibnamefont{Siewert}},
  \bibinfo{journal}{Sci.\ Rep.} \textbf{\bibinfo{volume}{2}},
  \bibinfo{pages}{942} (\bibinfo{year}{2012}).

\bibitem[{\citenamefont{Ma et~al.}(2011)\citenamefont{Ma, Chen, Chen, Spengler,
  Gabriel, and Huber}}]{Ma2011}
\bibinfo{author}{\bibfnamefont{Z.-H.} \bibnamefont{Ma}},
  \bibinfo{author}{\bibfnamefont{Z.-H.} \bibnamefont{Chen}},
  \bibinfo{author}{\bibfnamefont{J.-L.} \bibnamefont{Chen}},
  \bibinfo{author}{\bibfnamefont{C.}~\bibnamefont{Spengler}},
  \bibinfo{author}{\bibfnamefont{A.}~\bibnamefont{Gabriel}}, \bibnamefont{and}
  \bibinfo{author}{\bibfnamefont{M.}~\bibnamefont{Huber}},
  \bibinfo{journal}{Phys.\ Rev.\ A} \textbf{\bibinfo{volume}{83}},
  \bibinfo{pages}{062325} (\bibinfo{year}{2011}).

\bibitem[{\citenamefont{Vidal and Werner}(2002)}]{Vidal2002}
\bibinfo{author}{\bibfnamefont{G.}~\bibnamefont{Vidal}} \bibnamefont{and}
  \bibinfo{author}{\bibfnamefont{R.}~\bibnamefont{Werner}},
  \bibinfo{journal}{Phys.\ Rev.\ A} \textbf{\bibinfo{volume}{65}},
  \bibinfo{pages}{032314} (\bibinfo{year}{2002}).

\bibitem[{\citenamefont{Braunstein et~al.}(1999)\citenamefont{Braunstein,
  Caves, Jozsa, Linden, Popescu, and Schack}}]{BCJLPS1999}
\bibinfo{author}{\bibfnamefont{S.~L.} \bibnamefont{Braunstein}},
  \bibinfo{author}{\bibfnamefont{C.~M.} \bibnamefont{Caves}},
  \bibinfo{author}{\bibfnamefont{R.}~\bibnamefont{Jozsa}},
  \bibinfo{author}{\bibfnamefont{N.}~\bibnamefont{Linden}},
  \bibinfo{author}{\bibfnamefont{S.}~\bibnamefont{Popescu}}, \bibnamefont{and}
  \bibinfo{author}{\bibfnamefont{R.}~\bibnamefont{Schack}},
  \bibinfo{journal}{\prl} \textbf{\bibinfo{volume}{83}}, \bibinfo{pages}{1054}
  (\bibinfo{year}{1999}).

\end{thebibliography}


\end{document}